\documentclass{iopconfser}

\newcommand{\Xmax}{$X_{\mathrm{max}}$}

\usepackage{hyperref}
\usepackage{ragged2e}
\usepackage{graphicx}

\begin{document}

\title{Energy spectrum and mass composition of cosmic rays from Phase I data measured using the Pierre Auger Observatory}

\author{Vladim\'ir Novotn\'y$^{1}$ for the Pierre Auger Collaboration$^{2}$}

\affil{$^1$Institute of Particle and Nuclear Physics, Faculty of Mathematics and Physics, Charles~University, V Hole\v sovi\v ck\'ach 2, 180 00 Prague 8, Czech Republic}
\affil{$^2$Full author list: \href{http://www.auger.org/archive/authors_2024_06.html}{http://www.auger.org/archive/authors\_2024\_06.html}}

\email{spokespersons@auger.org}

\begin{abstract}
\justifying
The Pierre Auger Observatory concluded its first phase of data taking after seventeen years of operation.
The dataset collected by its surface and fluorescence detectors (FD and SD) provides us with the most precise estimates of the energy spectrum and mass composition of ultra-high energy cosmic rays yet available.
We present measurements of the depth of shower maximum, the main quantity used to derive species of primary particles, determined either from the direct observation of longitudinal profiles of showers by the FD, or indirectly through the analysis of signals in the SD stations.
The energy spectrum of primaries is also determined from both FD and SD measurements, where the former exhibits lower systematic uncertainty in the energy determination while the latter exploits unprecedentedly large exposure.
The data for primaries with energy below 1 EeV are also available thanks to the high-elevation telescopes of FD and the denser array of SD, making measurements possible down to 6 PeV and 60 PeV, respectively.
\end{abstract}

\section{Introduction}
\justifying
The Pierre Auger Observatory \cite{PierreAuger:2015eyc} collects data since 2004.
The first phase of data taking ended after 17 years of operation on $31^{\mathrm{st}}$ December 2020, now followed by the AugerPrime upgrade.
In Section~\ref{sec:spec} and \ref{sec:mass}, respectively, we present the Phase~I measurements of the energy spectrum and the mass composition of ultra-high energy cosmic rays (UHECRs), i.e. those primaries with energy exceeding $10^{18}\,\mathrm{eV} = 1\,\mathrm{EeV}$.
At such large energies, the flux of primaries is so small that we are restricted to the detection of secondary particles, the extensive air showers (EASs).
The data gathered at the highest energies using the surface and fluorescence detectors (FD and SD) are also complemented by lower-energy measurements making use of high-elevation telescopes (HEAT) and denser arrays of the SD with the spacing of 750\,m and 433\,m.

\section{Energy spectrum}
\label{sec:spec}
\justifying
The energy spectrum of UHECRs is measured using the SD, composed of water-Cherenkov stations separated by 1500\,m on a triangular grid covering an area of $\sim$3000\,km$^2$, and horizontally looking telescopes of the FD, placed at four sites around the SD, dominating the exposure and precision of the energy determination, respectively.
There exist two different reconstruction methods for the SD events incoming under the zenith angles below and above $60^{\circ}$ resulting in $\it vertical$ and $\it inclined$ datasets, respectively.
As described in detail in Refs.~\cite{spec1500_PhysRevD.102.062005,Novotný:2021sA}, the energy estimators from the SD are calibrated, using a common subset of high quality {\it hybrid} events, to the energies derived from the FD.
This procedure ensures that the common energy scale of the measurements is determined with systematic uncertainty of $14\%$, i.e. the systematics of the FD \cite{Dawson:2019pk}.
An equivalent approach is used for the data from the SD array with the spacing of 750\,m.

\begin{figure}[!h]
\centering
\includegraphics[width=117mm]{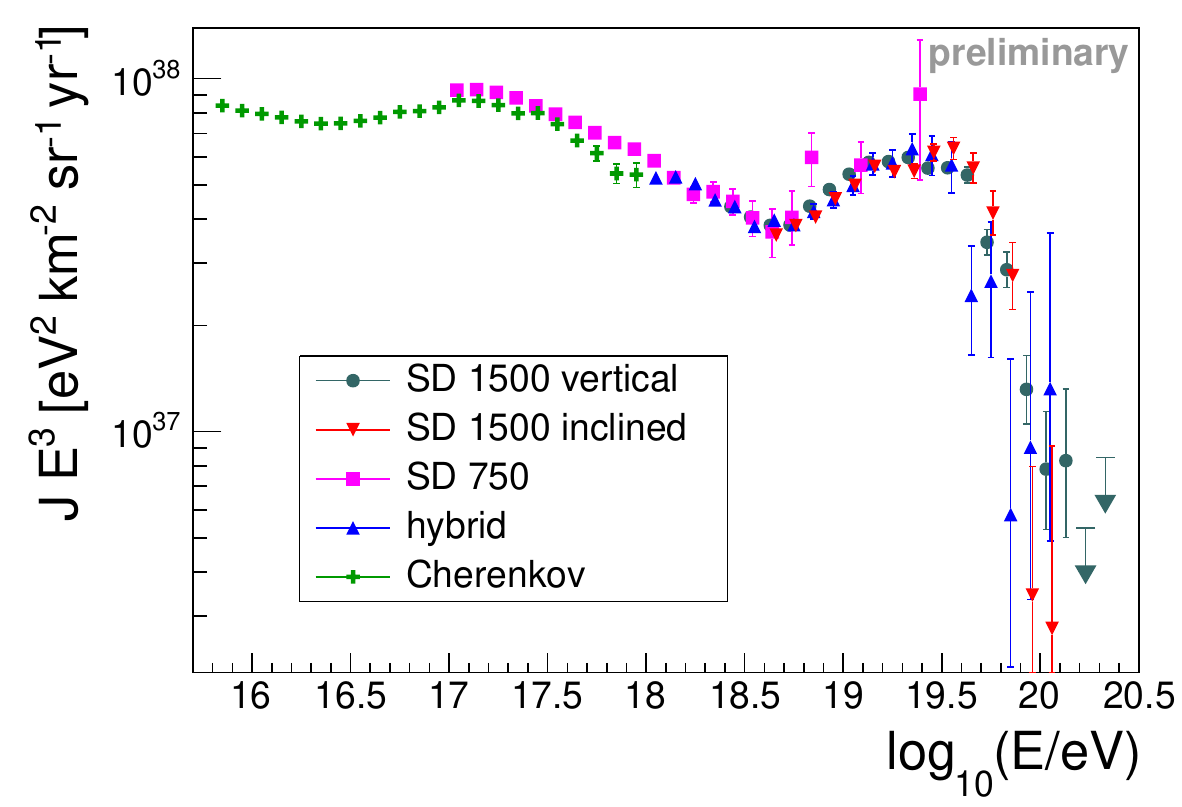}
\caption{
\label{fig:spectra}
Five estimates of the energy spectrum of cosmic rays derived from the Pierre Auger Observatory data using different methods described in the text.
Figure is from Ref.~\cite{Novotný:2021sA}.}
\end{figure}

The resulting energy spectra of the three SD methods are shown in Fig.~\ref{fig:spectra}, together with the one obtained from the {\it hybrid} dataset consisting of the FD-measured events, and the energy spectrum derived from the Cherenkov-dominated FD data \cite{Novotný:2021sA}.
The FD-based samples, although possessing better resolution of the energy determination, suffer from limited statistics associated with restricted time of operation to the clear moonless nights ($\sim$$14\%$ of the total time), while the SD operates constantly.
It is also important to emphasize that the exposure of the SD is geometrical, i.e. its aperture is formed by active hexagonal cells, while the exposure of the FD must be calculated from realistic Monte Carlo (MC) simulations.
This fact contributes to mutual uncertainties of the flux normalization between different measurements.

\begin{figure}[!h]
\centering
\includegraphics[width=117mm]{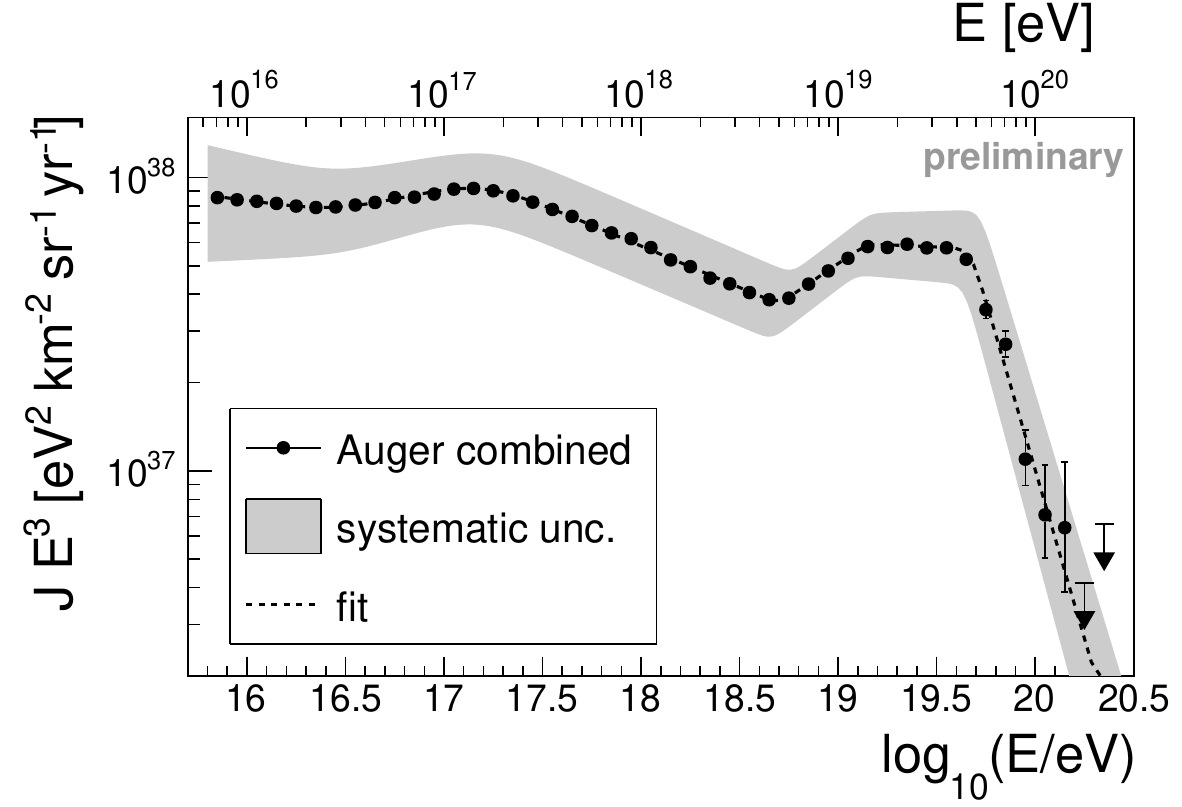}
\caption{
\label{fig:combined}
Energy spectrum combined from measurements shown in Fig.~\ref{fig:spectra}.
Changes in the intensity of cosmic rays are visually enhanced by multiplying by the third power of the energy.
The dashed line corresponds to the broken-power-law fit with smooth transitions.
Positions of the breaks are listed in the text.
Systematic uncertainty shown by the gray band is driven by the energy scale uncertainty of $14\%$.
Data come from Ref.~\cite{Novotný:2021sA}.}
\end{figure}

The five estimates of the energy spectrum are combined, taking into account residual systematic differences, into a single one depicted in Fig.~\ref{fig:combined}.
It covers the energy range from $10^{15.8}$\,eV up to the highest energies and reveals the {\it low-energy ankle} at $(2.8 \pm 0.3 \pm 0.4)\times 10^{16}$\,eV\footnote{Positions of breaks are given in the format (value $\pm$ stat. unc. $\pm$ syst. unc.).}, the {\it $2^{nd}$ knee} at $(1.58 \pm 0.05 \pm 0.2)\times 10^{17}$\,eV, the {\it ankle} at $(5.0 \pm 0.1 \pm 0.8) \times 10^{18}$\,eV, the {\it instep} starting at $(1.4 \pm 0.1 \pm 0.2)\times 10^{19}$\,eV, and the abrupt suppression above $(4.7 \pm 0.3 \pm 0.6)\times 10^{19}$\,eV.
Corresponding spectral indices ranges from 2.54 to 5.3 and can be found in Ref.~\cite{Novotný:2021sA}.

\begin{figure}[!h]
\centering
\includegraphics[width=117mm]{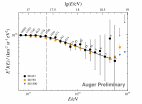}
\caption{
\label{fig:SDLE}
Energy spectrum of cosmic rays derived from the SD~433\,m measurement compared with the SD~750\,m and SD~1500\,m values
\cite{BrichettoOrquera:202340}.
Position of the {\it $2^{nd}$ knee} is marked by the vertical dashed line.
}
\end{figure}

The result shown in Fig.~\ref{fig:SDLE} from even denser SD array with the spacing of 433\,m, which is calibrated using the common data sample with the SD 750\,m array, recently confirmed the presence of the {\it $2^{nd}$ knee} at $(2.30 \pm 0.50 \pm 0.35)\times 10^{17}$\,eV by the robust SD method \cite{BrichettoOrquera:202340}.

\section{Mass composition}
\label{sec:mass}
To derive composition of UHECRs we restrict the search to stable nuclei and protons, because other primaries are extremely unlikely to reach the Earth with ultra-high energy\footnote{Searches for neutrinos, gammas and neutrons are targeted to specific scenarios and are beyond the scope of this contribution.}.
Motivated by relative abundances of nuclei in our Galaxy, the Fe is assumed to be the heaviest expected element among the UHECRs.

The essential quantities correlated with the mass of primary particle are the depth of the shower maximum, \Xmax, and the number of muons in the EAS.
In Phase~I of the Observatory measurements, we mostly rely on the determination of \Xmax, while the AugerPrime upgrade will incorporate the measurement of the muon content.
The most clear way of the \Xmax~determination is a direct detection of the longitudinal profile using the FD.
This method, while restricted in exposure and thus also in maximum probed energy, is still the most accurate one with the systematic uncertainty in the $\left< X_{\mathrm{max}} \right>$ determination $\leq10.2\,\mathrm{g}\,\mathrm{cm}^{-2}$ over the whole energy range \cite{xmaxold_PhysRevD.90.122005}.

An indirect \Xmax~measurement making use of the SD arrival times and measured signal traces, processed with the deep learning \cite{deep_AbdulHalim:2023C3}, is calibrated to the FD based \Xmax~estimates on a common subset of events.
This cross calibration involves a bias of $\sim$30$\,\mathrm{g}\,\mathrm{cm}^{-2}$ attributed to the mismatch between the real data and the EAS simulations, on which the neural network was trained.

The last available \Xmax~evaluation to date is derived from the Auger Engineering Radio Array (AERA), which consists of 153 radio antennas located within the grid of the SD on an area of 17\,km$^2$, close to one of the FD sites \cite{radio_PhysRevLett.132.021001}.
The \Xmax~reconstruction method uses detailed MC simulations of radio signals from EASs which are compared with those that are measured in AERA, taking the \Xmax~value from the closest simulation \cite{radio_PhysRevD.109.022002}.

\begin{figure}[!h]
\centering
\includegraphics[width=79mm]{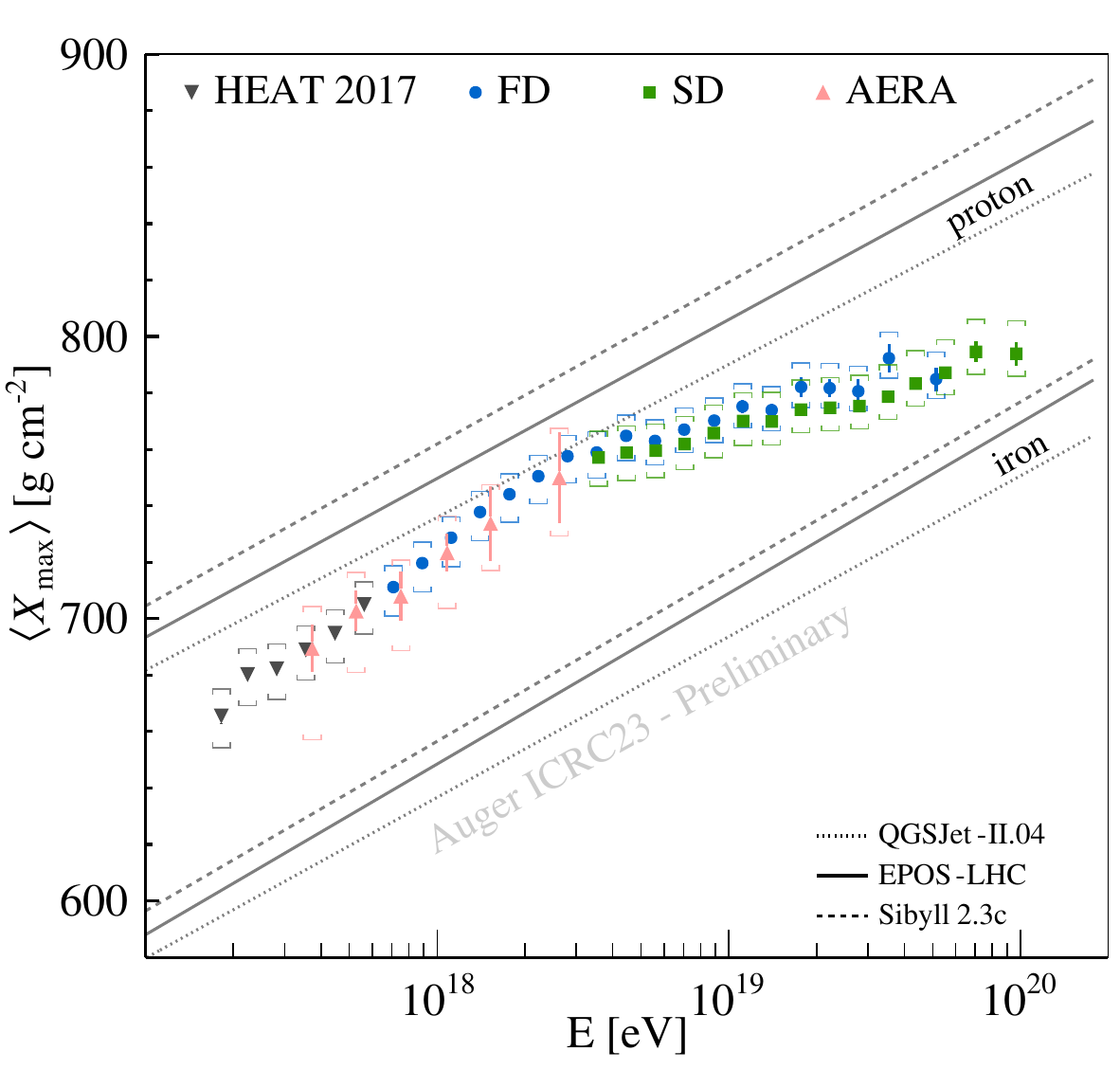}
\includegraphics[width=79mm]{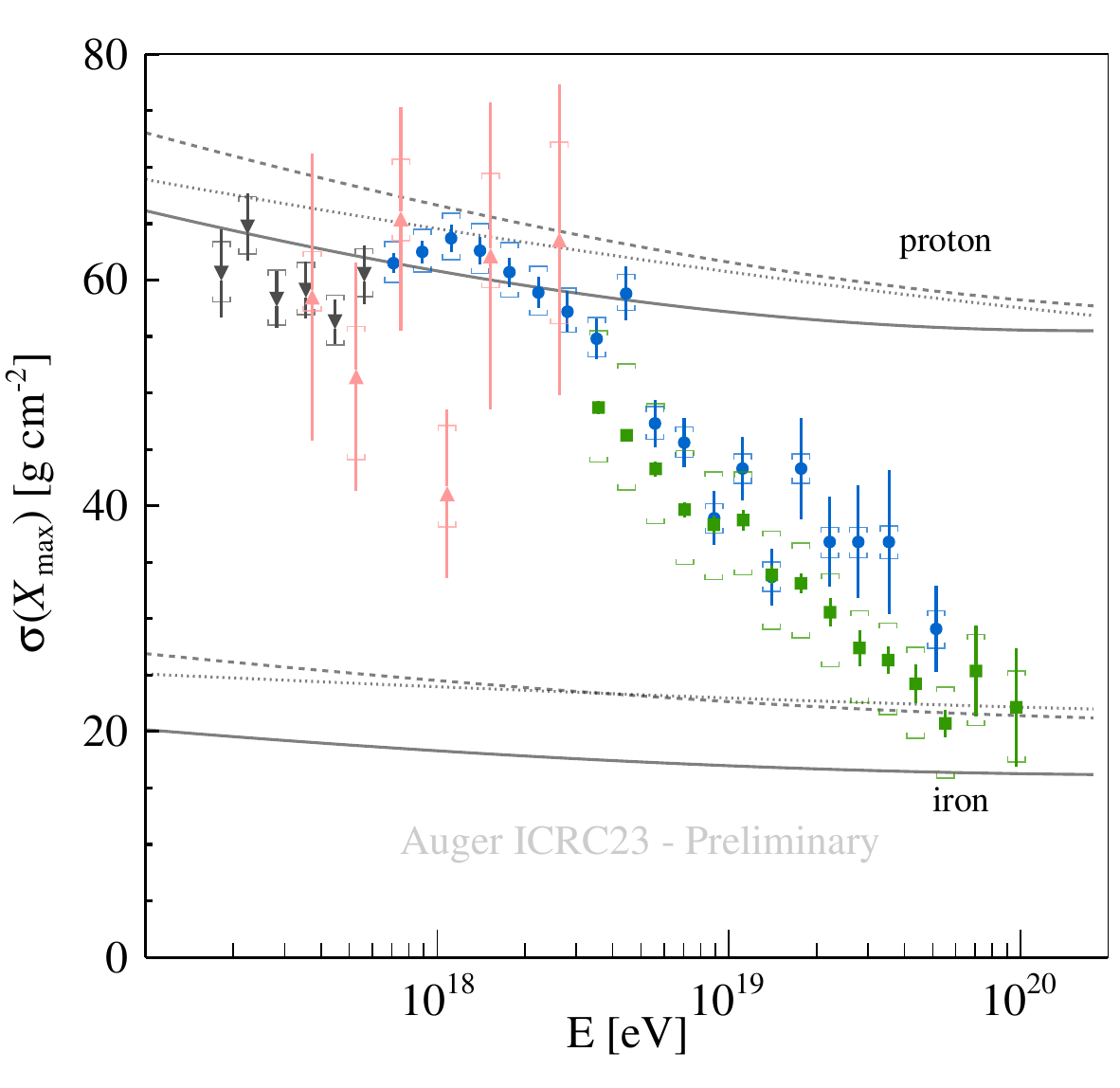}
\caption{
\label{fig:xmax}
Averages (left) and standard deviations (right) of measured \Xmax~distributions as functions of energy.
FD \cite{xmaxfd_AbdulHalim:20239/} corresponds to the measurements from horizontally looking telescopes of the FD,
HEAT~2017 \cite{Bellido:2017Li} marks the low-energy extension of the FD, SD is the estimate from deep learning \cite{deep_AbdulHalim:2023C3}, and
the radio dataset, which consists of only 594 showers, is marked as AERA \cite{radio_PhysRevLett.132.021001,radio_PhysRevD.109.022002}.
Reproduced from Ref.~\cite{Mayotte:2023Nc}
}
\end{figure}

All three above mentioned datasets are indicated in Fig.~\ref{fig:xmax}, where averages and standard deviations of the measured \Xmax~distributions are shown in the left and right panel, respectively.
The detector effects are unfolded.

\begin{figure}[!h]
\centering
\includegraphics[width=79mm]{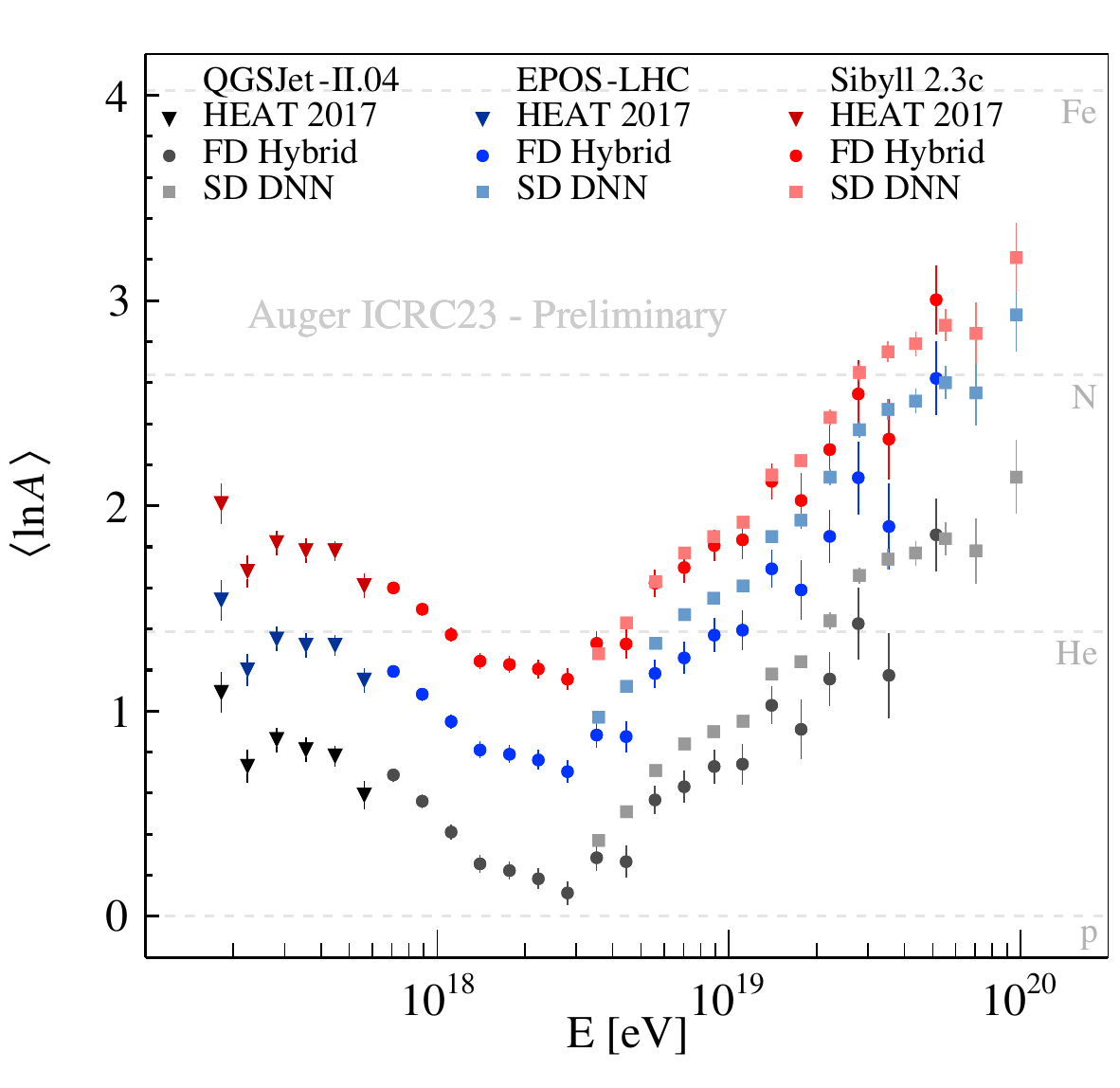}
\includegraphics[width=79mm]{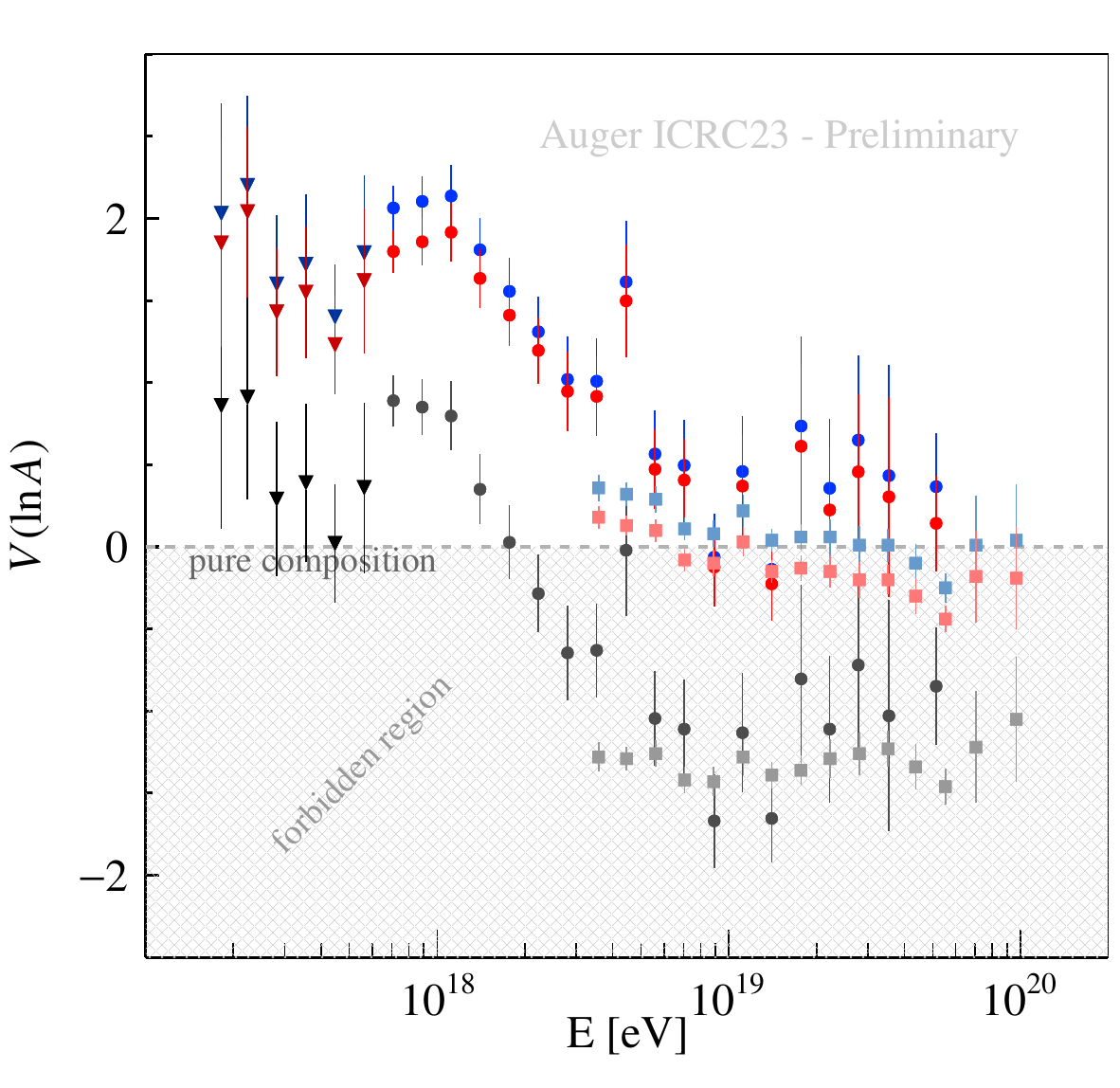}
\caption{
\label{fig:lnA}
Averages (left) and variances (right) of $\ln{A}$ inferred from \Xmax-moments using different high-energy interaction models.
Data correspond to those in Fig.~\ref{fig:xmax}. Comes from Ref.~\cite{Mayotte:2023Nc}.}
\vspace{-6mm}
\end{figure}

Using particular high-energy interaction model, the $\left< X_{\mathrm{max}} \right>$ and $\sigma^2 \left(X_{\mathrm{max}}\right)$ can be translated to the moments of the logarithmic mass, $\left< \ln{A} \right>$ and $\sigma^2 \left(\ln{A}\right)$, by inverting \cite{xmax_interpret}
\begin{eqnarray*}
    \left< X_{\mathrm{max}} \right> = \left< X_{\mathrm{max}} \right>_\mathrm{p} + f_E \left< \ln{A} \right>, \\
    \sigma^2 \left(X_{\mathrm{max}}\right) = \left<\sigma^2_{\mathrm{sh}}\right> + f^2_E\,\sigma^2\left(\ln{A} \right),
\end{eqnarray*}
where the model-dependent parameters are the mean \Xmax~for protons, $\left< X_{\mathrm{max}} \right>_\mathrm{p}$, the average shower-to-shower fluctuations, $\left<\sigma^2_{\mathrm{sh}}\right>$, and an energy-dependent parameter $f_E$.
The results of the $\ln{A}$ recalculation of \Xmax~are shown in Fig.~\ref{fig:lnA}, using QGSJetII-04 \cite{qgs_PhysRevD.81.114028}, EPOS-LHC \cite{epos_PhysRevC.92.034906}, and Sybill2.3c \cite{sib23c_Riehn:20171S} interaction models.
In the right panel of Fig.~\ref{fig:lnA}, where $V\left(\ln{A}\right)$ denotes $\sigma^2\left(\ln{A}\right)$, the filled region depicts an unphysical region of negative variances, stressing this way the non-compatibility between model predictions and measured data.

\begin{figure}[!b]
\centering
\includegraphics[width=130mm]{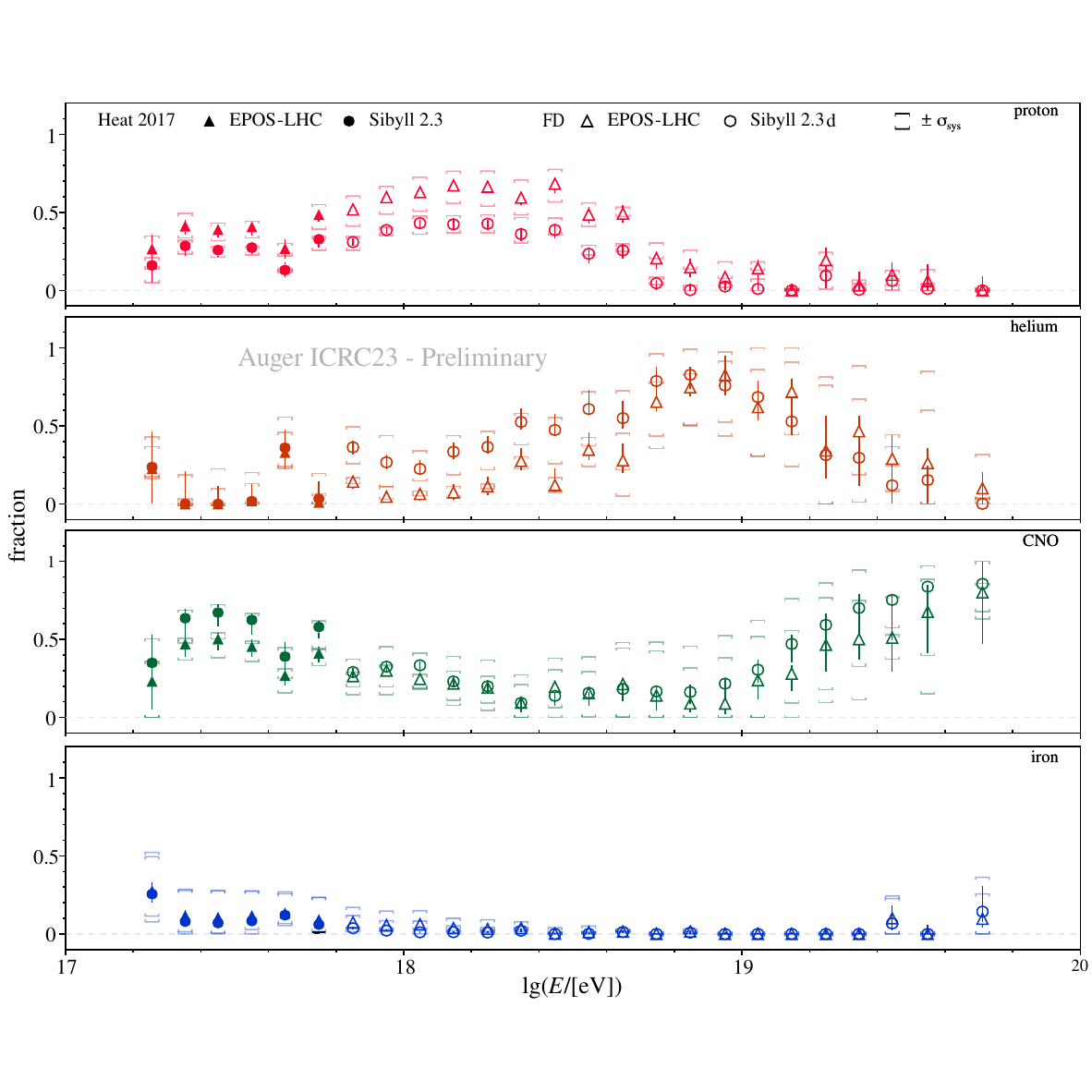}
\vspace{-10mm}
\caption{
\label{fig:fracs}
Elementary fractions as fitted to the FD \Xmax~distributions \cite{xmaxfd_AbdulHalim:20239/, Bellido:2017Li}.
Taken from Ref.~\cite{Mayotte:2023Nc}.}
\end{figure}

Using solely the direct measurements of \Xmax~from the FD, the \Xmax~distributions in each energy bin were fitted with MC predictions for individual linearly-in-mass-separated groups of primaries, namely modelled by protons, He, CNO, and Fe.
Resulting primary fractions are summarized in Fig.~\ref{fig:fracs}.
While subject of uncertainties in the models of hadronic interactions, the general trend of the composition evolving with energy is clearly visible and the pure proton scenario is ruled out.

The absolute scale of the model-predicted \Xmax~plays an important role in the interpretation of the mass composition data.
Recently, it was tested using the {\it hybrid} data measured independently by the FD and the SD \cite{vicha_PhysRevD.109.102001}.
By fitting the FD and SD signals simultaneously, it was possible to asses the level of disagreement between the model predictions and the measured data at the Pierre Auger Observatory.
It turns out that not only the modelled \Xmax~scale should be shifted by $\sim$30$\,\mathrm{g}\,\mathrm{cm}^{-2}$, but also the modelled muon content is too low by $\sim$$18\%$.
Both values are model-dependent and valid in the energy region around the ankle, where the analysis in Ref.~\cite{vicha_PhysRevD.109.102001} was performed.
These findings add up to the uncertainty in the inference of the mass composition and complete the previous inconsistencies of models shown in Fig.~\ref{fig:lnA}.

\section{Conclusions}
\justifying
In its Phase I, the Pierre Auger Observatory successfully measured, using several techniques, basic characteristics of UHECRs, namely their energy spectrum and the mass composition.
The energy spectrum clearly exhibits features colloquially named the {\it low-energy ankle}, the {\it $2^{nd}$ knee}, the {\it ankle}, the {\it instep} and a steep suppression above 47~EeV.
The mass composition seems to evolve according to Peters' cycle \cite{Peters:1961mxb}, being dominated by protons around 1 EeV, followed by helium nuclei around 10 EeV and the CNO group at about 50 EeV and above.
Nevertheless, this inference heavily depends on predictions of high-energy interaction models and will be precised with our knowledge of these interactions.

\section*{Acknowledgements}
This work was co-funded by the European Union and supported by the Czech Ministry of Education, Youth and Sports (Project No. FORTE - CZ.02.01.01/00/22\_008/0004632).

\bibliographystyle{iopart-num}
\bibliography{spectrum_mass}

\providecommand{\newblock}{}
\begin{thebibliography}{10}
\expandafter\ifx\csname url\endcsname\relax
  \def\url#1{{\tt #1}}\fi
\expandafter\ifx\csname urlprefix\endcsname\relax\def\urlprefix{URL }\fi
\providecommand{\eprint}[2][]{\url{#2}}

\bibitem{PierreAuger:2015eyc}
Aab A {\em et~al.\/} (The Pierre Auger Collaboration) 2015 {\em Nucl. Instrum. Meth. A\/} {\bf 798} 172--213

\bibitem{spec1500_PhysRevD.102.062005}
Aab A {\em et~al.\/} (The Pierre Auger Collaboration) 2020 {\em Phys. Rev. D\/} {\bf 102}(6) 062005

\bibitem{Novotný:2021sA}
Novotný V {\em et~al.\/} (The Pierre Auger Collaboration) 2021 {\em PoS\/} {\bf ICRC2021} 324

\bibitem{Dawson:2019pk}
Dawson B {\em et~al.\/} (The Pierre Auger Collaboration) 2019 {\em PoS\/} {\bf ICRC2019} 231

\bibitem{BrichettoOrquera:202340}
Brichetto~Orquera G {\em et~al.\/} (The Pierre Auger Collaboration) 2023 {\em PoS\/} {\bf ICRC2023} 398

\bibitem{xmaxold_PhysRevD.90.122005}
Aab A {\em et~al.\/} (The Pierre Auger Collaboration) 2014 {\em Phys. Rev. D\/} {\bf 90}(12) 122005

\bibitem{deep_AbdulHalim:2023C3}
Glombitza J {\em et~al.\/} (The Pierre Auger Collaboration) 2023 {\em PoS\/} {\bf ICRC2023} 278

\bibitem{radio_PhysRevLett.132.021001}
Abdul~Halim A {\em et~al.\/} (The Pierre Auger Collaboration) 2024 {\em Phys. Rev. Lett.\/} {\bf 132}(2) 021001

\bibitem{radio_PhysRevD.109.022002}
Abdul~Halim A {\em et~al.\/} (The Pierre Auger Collaboration) 2024 {\em Phys. Rev. D\/} {\bf 109}(2) 022002

\bibitem{xmaxfd_AbdulHalim:20239/}
Fitoussi T {\em et~al.\/} (The Pierre Auger Collaboration) 2023 {\em PoS\/} {\bf ICRC2023} 319

\bibitem{Bellido:2017Li}
Bellido J {\em et~al.\/} (The Pierre Auger Collaboration) 2017 {\em PoS\/} {\bf ICRC2017} 506

\bibitem{Mayotte:2023Nc}
Mayotte E~W {\em et~al.\/} (The Pierre Auger Collaboration) 2023 {\em PoS\/} {\bf ICRC2023} 365

\bibitem{xmax_interpret}
Abreu P {\em et~al.\/} (The Pierre Auger Collaboration) 2013 {\em JCAP\/} {\bf 2013} 026

\bibitem{qgs_PhysRevD.81.114028}
Ostapchenko S 2010 {\em Phys. Rev. D\/} {\bf 81}(11) 114028

\bibitem{epos_PhysRevC.92.034906}
Pierog T, Karpenko I, Katzy J~M, Yatsenko E and Werner K 2015 {\em Phys. Rev. C\/} {\bf 92}(3) 034906

\bibitem{sib23c_Riehn:20171S}
Riehn F, Dembinski H, Fedynitch A, Engel R, Gaisser T and Stanev T 2017 {\em PoS\/} {\bf ICRC2017} 301

\bibitem{vicha_PhysRevD.109.102001}
Abdul~Halim A {\em et~al.\/} (The Pierre Auger Collaboration) 2024 {\em Phys. Rev. D\/} {\bf 109}(10) 102001

\bibitem{Peters:1961mxb}
Peters B 1961 {\em Nuovo Cim.\/} {\bf 22} 800--819

\end{thebibliography}

\end{document}